# A new martian crater chronology: Implications for Jezero crater

Simone Marchi

Southwest Research Institute, 1050 Walnut St, Boulder, CO 80302, USA
marchi@boulder.swri.edu

**Abstract**

Crater chronologies are a fundamental tool to assess relative and absolute ages of planetary surfaces when direct radiometric dating is not available (e.g., Neukum et al 2001; Marchi et al 2009). Martian crater chronologies are derived from lunar crater spatial densities on terrains with known radiometric ages, and thus they critically depend on the extrapolation Moon to Mars (Ivanov 2001; Neukum et al 2001; Hartmann 2005). This extrapolation requires knowledge of the time evolution of the impact flux, including contributions from various impactor populations, factors that are not trivially connected to the dynamical evolution of the early Solar System.

In this paper, we will present a new martian crater chronology based on current dynamical models, and consider the main sources of uncertainties (e.g., impactor size-frequency distribution; dynamical models with late and early instabilities, etc). The resulting "envelope" of martian crater chronologies significantly differs from previous chronologies. The new martian crater chronology is discussed using two interesting applications: Jezero crater's dark terrain (relevant to the NASA Mars 2020 mission) and the southern heavily cratered highlands. Our results indicate that Jezero's dark terrain may have formed ~ 3.1 Ga, that is up to 0.5 Gyr older than previously thought (e.g., Shahrzad et al 2019). In addition, available crater chronologies (including our own) overestimate the number of craters larger than 150 km on the southern highlands, suggesting that either large craters have been efficiently erased over martian history, or that dynamical models need further refinement. Further, our chronology constrains the age of Isidis basin to be 4.05-4.2 Ga and Borealis basin to be 4.35-4.40 Ga, these are predictions that can be tested with future sample and return missions.

1. Introduction

The combination of radiometric ages of terrains sampled by Apollo and Luna missions have been widely used to derive lunar chronology functions, that is, number of craters per unit surface as a function of time. The lunar crater chronology is then extrapolated to Mars. Two main martian chronologies have been proposed to date (Neukum et al 2001; Hartmann 2005; and references therein). The main aspect of interest is that both chronologies rely on similarly simplistic assumptions about the evolution of the impactor population over time, and as such, these chronologies do not represent truly independent approaches, and their differences are mostly due to specific assumptions, such as the slopes of crater size frequency distributions. A more recent martian chronology sought to radiometrically calibrate the martian impact flux using Mojave crater as the source of the shergottite meteorites of known radiometric ages (Werner et al. 2014), and in-situ radiometric ages at Gale crater by the NASA *Curiosity* rover (Werner 2019). This chronology represents a novel approach as it aims to incorporate martian radiometric data into the martian chronology; however, the association of specific meteorite ages or in-situ radiometric ages with ancient landforms and crater counts remain uncertain.

Our understanding of the dynamical evolution of the Solar System and the time evolution of lunar and martian impact rates have greatly improved in recent years. In particular, current martian crater chronologies (e.g., Hartmann 2005; Werner 2019; see also Ivanov 2001 for details) rely on Moon to Mars extrapolations based on present impact rates estimated from near-Earth objects, and it has not been addressed whether and to what extent they are compatible with the dynamical evolution of the inner early Solar System. In this paper, we propose a revised martian crater chronology compatible with current understanding of the sources of impactors in the terrestrial planet region, and their evolution over the history of the Solar System, focusing on the earliest ($\geq 2$ Ga) impact flux.

## 2. Summary of lunar impact flux over time

We start by presenting how lunar data can be used to calibrate its cratering history. Solar System dynamical models show that there are various populations of lunar impactors, namely asteroids and leftover planetesimals from the terrestrial planet region, and comets. The relative proportions of these impactor populations and their temporal evolutions, however, are strongly dependent on the specific dynamical evolution of the Solar System. There is widespread agreement among dynamicians (e.g., Nesvorny 2018 for a review) that the giant planets may have formed in a compact configuration, which due to ensuing orbital migration led to an unstable configuration (e.g., Tsiganis et al 2005). Such a planetary instability is needed to explain a number of key features of the current architecture of the Solar System, including planetary eccentricity and inclinations, orbital excitation of the Kuiper Belt and Main Belt, etc. Recent dynamical models favor an instability within the first 100 Myr of so after formation (Nesvorny et al 2018; Clement et al 2018), but here we present two end member scenarios encompassing the variability in proposed models: we consider an early instability (e.g., Nesvorny et al 2017) and a late instability model at 4.1 Ga (e.g., Marchi et al 2013).

*Early instability model.* Support for this scenario comes from recent dynamical simulations of the early Solar System. It has been shown that an early instability (within 100 Myr since formation) better explains the orbital architecture of the terrestrial planets (Agnor and Lin 2012), as well as Main Belt asteroids (MBA; Deienno et al 2017; Clement et al 2018). Morbidelli et al (2018) showed that a combination of leftover planetesimals and asteroids leaving the Main Belt could explain lunar cratering data with an instability at ~4.5 Ga (Fig. 1a). A key feature of this model is that most lunar craters larger than 1 km in diameter ($N_1$) are produced by leftover planetesimals. Here we used the asteroid flux by Nesvorny et al (2017) for an early instability that has been calibrated using the number of current MBAs larger than 10 km. The resulting lunar asteroid impactor flux is two orders of magnitudes lower than leftover planetesimals. Notably, to achieve a good fit for ages ~3.5-3.7 Ga, Morbidelli et al (2018) included a constant flux of impactors resulting in a linear cumulative number of craters. This is motivated by the production of lunar impactors from small asteroids removed from the Main Belt via non-gravitational forces as indicated by the short-lived near-Earth objects (e.g., Gravnik et al 2018). Finally, we have here neglected cometary impacts because they do not contribute significantly to the overall crater chronology, except for a 10-20 Myr spike at 4.5 Ga that is not resolvable in cratering study. We note that the lack of cometary-like material in lunar rocks (Joy et al 2012) may be compatible with such an early delivery, which may not have allowed this material to be preserved.

*Late instability model.* Morbidelli et al (2018) also presented a late instability scenario. They argued that the best fit was achieved for an instability at 3.9 Ga. Here, we explore this possibility, but argue that if there was a late instability, this was at 4.1 Ga and not 3.9 Ga. This is motivated by the Ar-Ar age of asteroidal meteorites (Marchi et al 2013) and possible terrestrial zircons (Marchi et al 2014). We find, in agreement with Morbidelli et al (2018), that the adopted asteroid impact flux (from Nesvorny et al 2017 for a late instability) does not allow to obtain a good fit for an instability at 4.1 Ga. However, Brasser et al (2020) argued that the early main belt asteroids could have been more numerous than previously considered, leading to a higher impact flux. We find that a good fit of lunar data and a 4.1 Ga instability can be obtained by increasing the asteroid flux by a factor of ~ 4, in line with Brasser et al (2020), and reducing the leftover planetesimals population to 30% of what assumed for an early instability case. As before, we added a constant impactor flux due to non-gravitational forces (Fig. 1b), but note that now the best fit of lunar $N_1$ data is achieved with a lower flux. This model implies that leftovers dominated the flux prior to ~ 4.3 Ga and that asteroids became the principal source of impactor afterward. As before, we neglect comets, although they could have provided a significant source of impacts within 10-20 Myr following the instability. This is motivated by the lack of evidence for a significant flux of cometary-like material in lunar rocks (Joy et al 2012), possibly due to the decimation of small pristine comets as they approached the inner solar system (e.g., Levison et al 2002). At any rate, our $N_1$ chronology should be considered a lower limit, if indeed the cometary flux was significant.

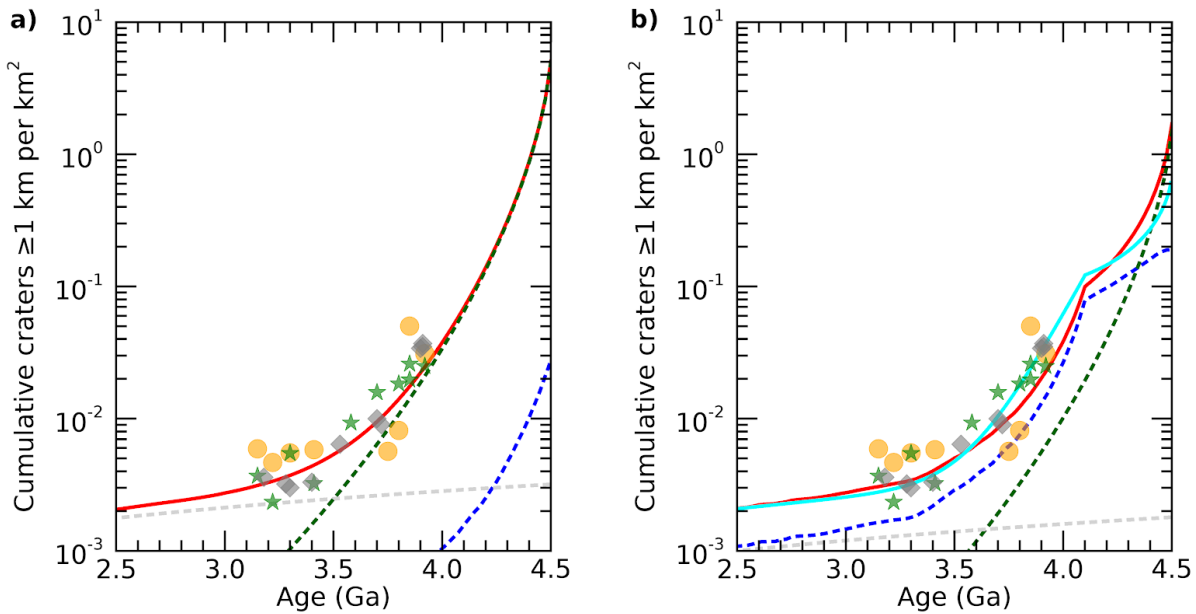

*Figure 1.* Lunar impact flux for early (a) and late instability (b). Symbols are for lunar $N_1$ values from various sources (Robbins 2014, orange circles; Marchi et al 2009, green stars; Neukum et al 2001, grey diamonds). a: The overall crater chronology (red) is the sum of leftover

*planetesimals (green), asteroids (blues), and constant flux (grey). The red curve is the same presented in Morbidelli et al (2018). b: Same notation as in panel a. For reference, the sawtooth model from Mobidelli et al (2012) is also indicated (cyan). See text for details.*

### 3. A new martian crater chronology

In order to properly extrapolate lunar chronology models to Mars we need to consider the nature of the impactors, as different populations may require different scaling. For instance, impactor populations may strike the Moon and Mars with different impact probabilities and velocities.

We adopt an average of results from the dynamical models from Raymond et al (2013), Nesvorny et al (2017), Morbidelli et al (2018), Brasser et al (2020). We note that the average impact velocities may vary by up to 10-15% across the dynamical models, and here we take an average impact velocity for asteroids of 23 and 13 km/s for the Moon and Mars, respectively. Moreover, we adopt an average impact velocity for leftover planetesimals of 13 km/s, for both Moon and Mars. Further, we consider that the ratio of martian-to-lunar impacts per unit surface is 2.6 and 0.5 (Nesvorny et al 2017; Morbidelli et al 2018), for asteroids and leftover planetesimals, respectively. These values do not strongly depend on the specific dynamical model. An additional complication is that smaller martian impactors are progressively eroded by atmospheric drag before reaching the ground. Here, we assume that the flux of projectiles resulting in 1 km martian crater are reduced by a factor 0.83 (Hartmann 2005).

The first step is to compute the impactor size ($d_1$) resulting in a 1 km crater. For this, we consider the cratering scaling laws from Melosh (1989), Holsapple and Housen (2007), and Johnson et al (2016). For the Moon, we find $d_1$ = 20-29 m and 29-41 m for asteroids and leftover planetesimals, respectively. The impactors size range for each case corresponds to different cratering scaling laws. We note that the Holsapple and Housen (2007)'s crater scaling law gives very similar results to the recently developed Johnson et al (2016)'s crater scaling law, thus we will adopt the latter in our nominal model. This corresponds to $d_1$ = 29 m and 41 m for asteroids and leftover planetesimals, respectively. Similar calculations for Mars produce $d_1$ = 34-42 m, for both asteroids and leftover planetesimals, with $d_1$ = 42 m is our nominal model. These values are only slightly depending on assumed physical parameters such as projectile and target density, and the range considered here is relatively large. For these calculations, we assumed a target and projectile densities of 2500 kg/m$^3$, target strength 2 MPa, impact angle of 45 deg.

Finally, we consider two characteristic impactor size frequency distributions (SFD), that is, near-Earth objects (NEO; Harris and D'Abramo 2015) and a recent model MBA model (Bottke et al 2020). Note that the impactor SFD does not have an important role for the lunar impact flux reported in Fig. 1. This is because the various impact fluxes have multiplicative factors that allow a similar fit of the $N_1$ data. For instance, the scaling factor used for the asteroid flux in Fig. 1b changes from about 4.5 to 2 for the MBA and NEO SFD, respectively.

We now have all the ingredients to extrapolate the lunar chronology to Mars. For this, we consider each lunar impactor population and compute the corresponding impact flux for Mars using relevant scaling parameters. We assume the asteroid flux and the constant flux have the same scaling factors as they both originate from MBAs. This may not be rigorously true, as these populations are produced by different dynamical processes, however, the differences are likely small compared with other sources of uncertainties (e.g., crater scaling laws). The results are shown in Fig. 2.

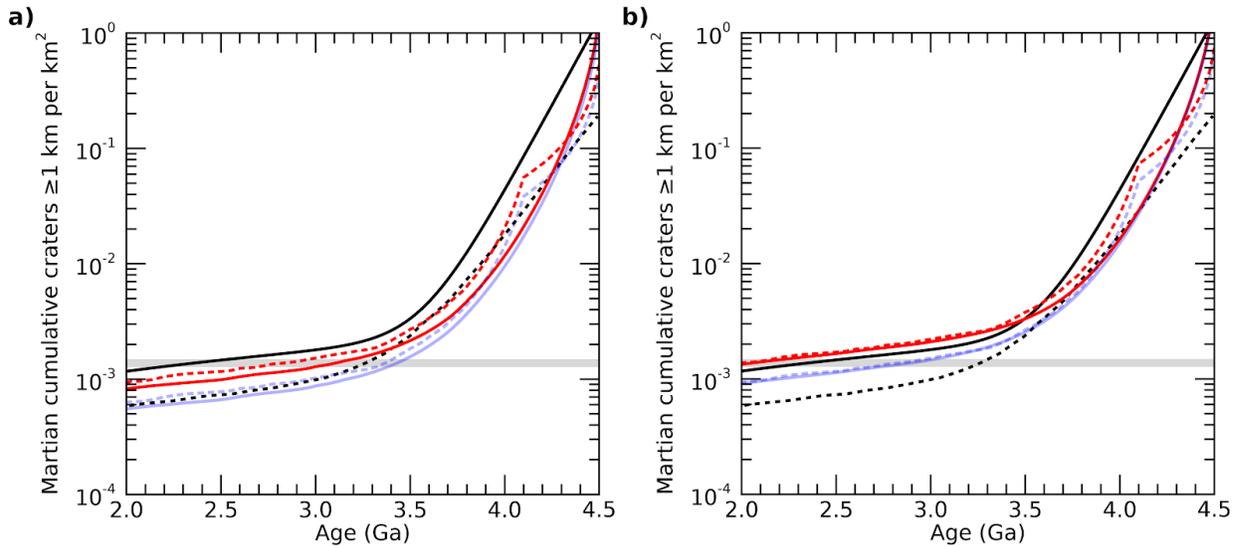

*Figure 2. Martian $N_1$ ($\geq 1$ km) cumulative crater distributions. The plots are for impactor sizes $d_1 = 34, 42$ m (a and b panels, respectively). Panel b is our nominal model. For each panel, dashed curves are for a late instability, and solid curves are for early instability. Red and blue indicate MBA and NEO SFDs, respectively. The Hartmann (2005) and Werner et al (2014) chronology curves are indicated (solid and dashed black curves, respectively). The gray lines indicate the range of $N_1$ values for Jezero crater (Shahrzad et al 2019). The new $N_1$ chronology curves for panel b) are provided in the Supplemental Material.*

An interesting aspect is that the extrapolation presented in Fig. 2 is specific for small impactors producing 1 km martian craters. It is commonly assumed that the $N_1$ crater chronology can be used to estimate the formation age of larger craters by simply rescaling for the corresponding number of impactors. This may not be correct, however, if multiple populations of impactors contribute to $N_1$, and if these populations require different scalings factors, for instance because the impact velocity are different, as discussed above for the Moon. In addition, going from $N_1$ craters to larger craters (thus large impactors) may require the suppression of the small impactor flux generated by non-gravitational forces, such as the Yarkovsky effect. We show this by computing the martian impact flux for projectiles $\geq 12$ km, which are responsible to make craters $\geq 150$ km ($N_{150}$ hereafter) according to the crater scaling laws discussed above. For this, we ignore the constant flux of small impactors as asteroids larger than ~ 5 km are not susceptible to non-gravitational forces. In addition, the extrapolation from the Moon to Mars of asteroids and leftover planetesimals require different scaling factors as these populations have different impact velocities. The results are presented in Fig. 3.

There are several things of interest in Fig. 3. First, the adopted impactor SFD has a significant effect on the early instability $N_{150}$ chronology. At 4.5 Ga, the MBA-based chronology is about a factor of ~ 4 higher than the NEO-based chronology. At 4.5 Ga the flux is entirely due to leftover planetesimals (Fig. 1a), and the factor of ~ 4 corresponds to the difference due to the MBA vs NEO SFDs in rescaling from 41 m (leftover planetesimal size responsible for 1 km craters on the Moon) to 12 km (responsible for 150 km craters on Mars). Note that our early instability chronologies differ from Morbidelli et al (2018). This is due to two main reasons. Morbidelli et al (2018) used an empirical factor of 1400 to convert 1 km lunar craters to 20 km lunar craters. This factor is from Neukum and Ivanov's production function (Neukum & Ivanov 1994). Then, they converted 20 km lunar craters to 150 km martian craters. Note that the empirical factor is obtained from a crater SFD that has been merged using crater counts from terrains of different ages and properties. Therefore this factor may be affected by systematic errors (e.g., changes in slope of the impactor SFD over time). We think it is more rigorous to remove this potential source of error by converting 1 km lunar crater directly to impactors, and then extrapolate to Mars. In addition, Morbidelli et al (2018) used different impact velocities for the Moon (18 km/s) and Mars (14 km/s), without making distinction of the different impactor populations (asteroids vs leftover planetesimals). To illustrate these differences, the Morbidelli et al (2018) chronology at 4.5 Ga is a factor of ~3 lower than our MBA-based chronology.

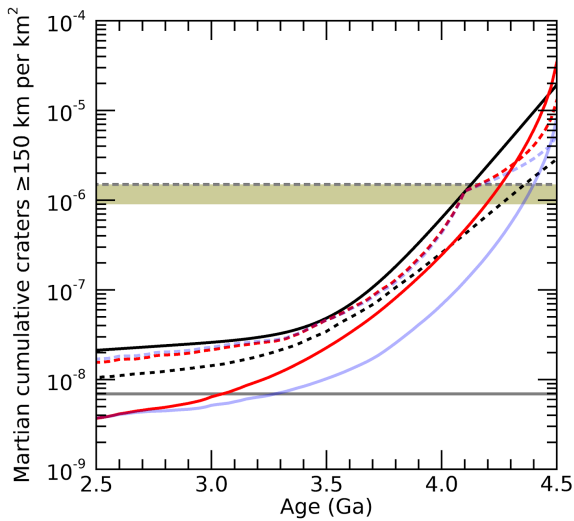

*Figure 3. Martian $N_{150}$ ($\geq$ 150 km) cumulative crater distribution. The plot is derived for impactors $\geq$ 12 km (for our nominal model, i.e., using Johnson et al 2016 cratering scaling law). Solid and dashed curves are for an early and late instability, respectively. Red and blue indicate MBA and NEO SFDs, respectively. The horizontal dashed grey line indicates the spatial density of 150 km on Mars's highlands, while the horizontal solid line corresponds to one crater over the entire surface. The solid and dashed black curves are rescaled versions of the $N_1$ chronology from Hartmann (2005) and Werner et al (2014), respectively (see text). The horizontal green area indicates the range of $N_{150}$ craters superposed on Isidis basin. The new $N_{150}$ chronology curves are provided in the Supplemental Material.*

For the late instability chronology the differences due to the impactor SFDs are less pronounced. This is because asteroids are the primary contribution to the impact flux, and here we compute the impact flux of 12 km impactors rescaling from the 10 km impactors (the latter are from Nesvorny et al 2017), without having to go through 1 km lunar craters.

For completeness, we report in Figure 3 approximate $N_{150}$ chronologies obtained by rescaling $N_1$ chronologies from Hartmann (2005) and Werner et al (2014). In the literature, crater size rescaling for the chronology functions is commonly done by using a crater production function or, equivalently, an impactor SFD. Accordingly, the rescaling is done here using the MBA SFD from 42 m ($N_1$) to 12 km ($N_{150}$). We stress, however, that this approach neglects the role of non-gravitational effects, as discussed before, and therefore we do not expect the result to be accurate.

## 4. Monte Carlo simulations

The crater chronologies presented in the previous section can be used to compute the number of martian craters ≥ 1 and 150 km as a function of time. However, the number of large basins are better estimated via a Monte Carlo approach due to the low number statistics. Thus, following the approach by Marchi et al (2014), we run Monte Carlo simulations that randomly generate projectiles ≥ 12 km (corresponding to 150 km craters). For this, we consider both crater chronologies derived for early and late instabilities and MBA-like SFD (Fig. 3, red curves). The MBA-like SFD has been extended to planetesimals 4000 km in diameter, as done in Marchi et al (2014). The results of the Monte Carlo simulations are shown in Fig. 4.

Our simulations show that cratering was a prominent process in early Mars evolution, with ~50-100% of the surface directly covered by craters ≥ 150 km in diameter (and many more smaller craters should have taken place too). In addition, we analysed the timing of impactors ≥ 500 km in diameter that strike Mars. A 500 km impactor, with a density of 3000 kg/m$^3$ and impact velocity of 13 km/s has an energy of ~2 x 10$^{28}$ J, that is the lower limit for Borealis-scale events (Marinova et al 2011; Nimmo et al 2008). The average impact time is ~4.42 Ga for both an early and late instability. Interestingly, the minimum time for an impactor ≥ 500 km to strike Mars is 4.3-4.35 Ga in about 15-20% of the simulations for a late and early instability, respectively. In addition, we find that 50 to 85% of the simulations have at least one impactor larger than 500 km for late and early instability, respectively. In these simulations, we find that for a late [early] instability the number of large impactors ranges from 1 to 2 [1 to 3]. Thus, the statistics of the Borealis-scale events does not show preference for a late or early instability, and both models are compatible with a recent analysis of the collisional delivery of martian highly siderophile elements (Marchi et al 2020). We will further discuss the timing of Borealis-scale events in the next section.

## 5. Discussion

The new martian $N_1$ crater chronology significantly differs from current chronologies (e.g., Hartmann 2005; Werner 2019). Uncertainties in cratering scaling laws and impactor SFDs (previously not fully taken into account for) play a significant role. Here we will discuss the implications of the new chronology with two examples: Jezero crater (the landing site of the NASA Mars 2020 mission), and the southern highlands.

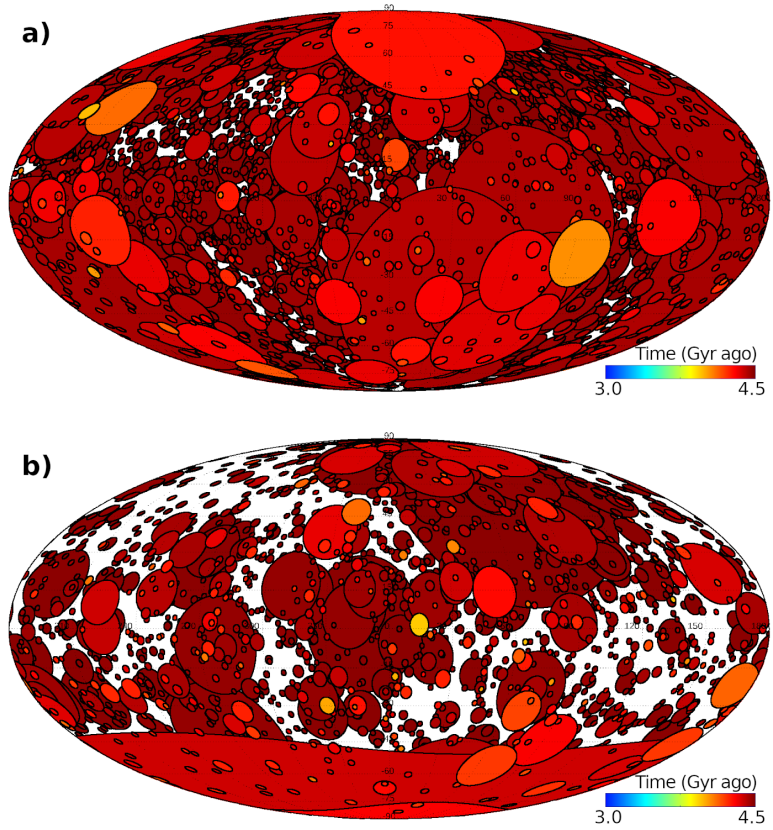

***Figure 4.*** *Monte Carlo simulations of Mars' bombardment history for an early (a) and late (b) instability. Circles have diameters 10x the impactor size, an indicative scale for crater sizes. Because final crater size is a function of the crustal temperature (Miljkovic et al 2013), we did not attempt to accurately predict crater sizes. Color coding indicates time of formation. Note the presence of large-scale collisions (projectiles ≥ 500 km) possibly compatible with Borealis-scale events for both scenarios (see text).*

For Jezero crater, we rely on crater counts from a dark-toned, mafic unit within the crater from Shahrzad et al (2019). This unit was emplaced after the formation of Jezero crater itself, and possibly post-dating water-related activity within the crater. For the following analysis, we use our nominal model based on Johnson et al (2016) crater scaling law and the NEO-like SFD. With these assumptions, the age of Jezero dark-toned unit is 2.7-3.1 Ga, depending on the uncertainties in the crater $N_1$ resulting from using small vs large craters (Shahrzad et al 2019). The impactor SFD has important implications for the derived dark-toned unit ages, and a MBA-like SFD would result in a younger age 1.9-2.3 Ga. Here we prefer using the NEO-like SFD following Strom et al (2005) and Holo and Kite (2020). Previous age estimates for the dark-toned units are ~ 1.4 Ga (Schon et al 2012), 2.6±0.2 Ga (Shahrzad et al 2019), and $3.35\pm^{0.16}_{-1.06}$ (Shahrzad et al 2019; the latter age is derived from Goudge et al 2012 and corrected to use Hartmann (2005) chronology, as used in the other age estimates). Thus, the difference in age estimates is solely due to uncertainties in crater counts because they all use the same chronology model. The most reliable nominal age of ~ 2.6 Ga (Shahrzad et al 2019; small craters, $N_1$ = 1.49x10$^{-3}$ km$^{-2}$) corresponds to our ~ 3.1 Ga, thus a net difference of about 0.5 Ga with respect to the Hartmann (2005) chronology. We estimate that the corresponding age for the Werner et al (2014) chronology is ~ 3.3 Ga. Complicating the issue further, a recent crater SFD for the Jezero dark-toned, mafic unit returned $N_1$ = 1.1x10$^{-3}$ km$^{-2}$ (Warner et al 2020), resulting in a ~ 2.45 Ga age in our nominal chronology.

Jezero crater is located near the NW portion of the rim of the 1500-km diameter Isidis basin. Thus, it seems possible that rocks near Jezero crater to be collected by the Mars 2020 mission may retain a radiometric age of the Isidis basin. For instance, Mustard et al (2009) argued that the presence of olivine-rich units near Jezero crater may be the result of Isidis impact melting, and if so, rocks from these units are likely to return Isidis basin age. For this reason, we compute the Isidis basin age using superposed craters and our new chronology. The Isidis basin has a spatial density of superposed craters ≥ 50 km ($N_{50}$) of approximately 29±8 x 10$^{-6}$ km$^{-2}$ (Robbins et al 2013; Fassett and Head 2011). Craters of 50 km in diameter are made by ~ 4 km impactors. At this size, impactor's orbits are likely little affected by non-gravitational effects, so we can use the $N_{150}$ chronology. Using the MBA-like SFD and the nominal crater scaling law, we obtain $N_{50}/N_{150}$ = 23.9, thus Isidis basin $N_{150}$ = 1.2±0.3 x 10$^{-6}$ km$^{-2}$. The corresponding Isidis basin age ranges from 4.05 to 4.20 Ga for a late and early instability, respectively (using the MBA-based chronology). Previous Isidis basin age estimates range from 3.86 to 4.04 Ga (Robbins et al 2013). Note that the early instability age is significantly older than any of the previous estimates (see Conclusions).

Further, we use the new crater chronology to assess the number of craters ≥ 150 km that formed throughout martian history. The martian southern highlands indicate that $N_{150}$ ~ 1.5 x 10$^{-6}$ per km$^2$ (Morbidelli et al 2018; Fig. 5), and we stress that value based on observed craters is a lower limit for the true number of 150 km craters that ever formed on Mars. Our new chronology systematically predicts more $N_{150}$ craters than observed, regardless of the impactor SFD and timing of instability. For a late instability the excess of 150 km craters ranges from a factor of ~ 2.5 (NEO-like SFD) to ~ 9 (MBA-like SFD). For an early instability, the excess ranges from a factor of ~ 9 (NEO-like SFD) to ~ 20 (MBA-like SFD). Thus, assuming that a MBA-like SFD may be a better fit for ancient crater SFDs (e.g., Strom et al 2005), then the new crater chorology predicts from about 9 to 20 more $N_{150}$ craters formed on Mars than currently observed. Morbidelli et al (2018) argued that the discrepancy between predicted and observed $N_{150}$ craters can be reconciled by the global crater erasing due to the formation of the Borealis basin. Numerical simulations of collisions corresponding to a Borealis-scale event (e.g., Marinova et al 2011) indicate significant global surface damage, but the process of crater erasing due to Borealis formation has not been investigated in detail. Assuming global $N_{150}$ crater erasing does take place, then the $N_{150}$ constraint implies an age for Borealis basin age ranging from 4.2 Ga (late instability; early instability MBA-like SFD) to 4.4 Ga (early instability, NEO-like SFD). Thus, assuming that an impactor population at collisional equilibrium with a MBA-like SFD is more probable for early large impactors (up to Ceres-sized ~1000 km in diameter), then Borealis basin needs to be relatively young (~ 4.2 Ga). This is 100-150 Ma younger than predicted by our Monte Carlo simulations, as discussed in the previous section, thus indicating a disconnect between the impact flux age of Borealis basin and its required age by the $N_{150}$ constraint.

We point out that the formation of the Borealis basin may not have been the only process responsible for large scale crater erasure. For instance, Manske et al (2020) showed that large craters (≥ 150 km) that formed on a relatively warm, early Mars would have generated impact melt volumes in excess of their transient cavities. Under these conditions, it is likely that no topographic expression of these craters is preserved. In addition, geological evidence indicates that many large and old martian craters have been heavily eroded by aeolian, sedimentary and fluvial processes (e.g., Robbins and Hynek 2012).

Thus, a late formation of the Borealis basin coupled with a non-preservation of post-basin large craters could be compatible with the proposed martian crater chronology. Note that the rescaled Hartmann (2005) chronology (assuming a MBA-like SFD) results in a higher number of early impacts, so requiring an even younger Borealis basin and/or more large craters to be erased. For instance, Werner (2014) indicates that the oldest basin on Mars may be ~ 4.1 Ga, implying that a longer hiatus may exist in Mars cratering record. Interestingly, Werner et al (2014) predict significantly fewer large craters, but because of a shallower slope, the derived age of Borealis basin is ~ 4.35 Ga, that is similar to our estimates.

In addition, Bottke and Andrews-Hanna (2017) noted that the early bombardment history of Mars could be further constrained by geophysical properties of the Borealis basin rim. Mars has 4 unambiguous basins ≥ 500 km in diameter that formed post-Borealis (Hellas, Isidis, and Argyre, Utopia). They indicated that at most 12 basins ≥ 500 km could have formed after the Borealis basin. A larger number would have caused damage to the Borealis' rims contrary to observations. Thus, increasing $N_{150}$ by the same amount (a factor of ~ 3), the resulting inferred age of Borealis basin would be 4.35-4.40 Ga (early/late instability, MBA-like SFD; Figure 3). This age estimate is compatible with our Monte Carlo analysis, and may be compatible also with the observed $N_{150}$, if some of the large craters were erased post Borealis formation, as noted above.

In conclusion, the martian record of $N_{150}$ craters is at odds with current dynamical models (or any previous chronologies), suggesting that either Borealis basin formed late (4.20 to 4.35 Ga) and there was complete global erasure of preexisting large craters; or it formed 4.35-4.40 Ga and still 50% of the $N_{150}$ craters that formed post Borealis basin were erased. We find that with available data it is not possible to come to a firm conclusion whether a late or early instability model works better. Either case requires the erasing of a significant fraction of large craters post Borealis basin. Alternatively, the earliest impact flux models are not correct, for instance, requiring a steeper slope, or an impactor SFD with fewer objects ≥ 12 km.

To better assess the latter point, we computed the predicted crater SFD resulting from an impactor MBA-like SFD and assumed an impact velocity of 13 km/s. The derived crater model production function (MPF; Marchi et al 2009) is shown in Fig. 5. The MPF closely resembles the observed crater SFD for the martian highland in the crater size range 150-600 km. However, it underestimates the number of craters larger than 1000 km by a factor of ~ 2. Note that this could simply be due to low number statistics as there are only 4 craters ≥ 1000 km on the highlands. We warn, however, that the crater scaling law is largely uncertain at these basin sizes. For instance, the crust temperature could affect basin size by some ~ 20-40% (Miljkovic et al 2013). If the four large basins were to have an increased size for these thermal effects, then there will be no significant discrepancy between MPF and cratering data. Alternatively, the number of 150-600 km craters could have been reduced by erosion, as discussed earlier. In conclusion, we find no evidence that a non-MBA-like SFD may provide a better fit the highlands crater SFD.

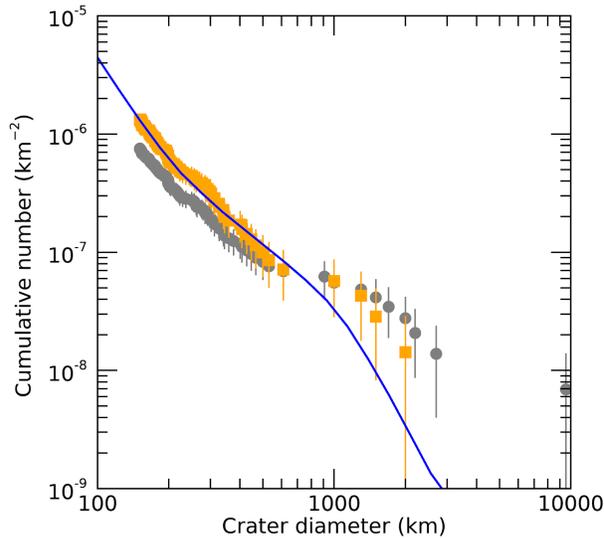

***Figure 5.** Martian craters ≥ 150 km SFD for the entire surface (gray; including Borealis basin), and highlands (orange). The latter contains 4 basins larger than 1000 km, that is, Hellas (2000 km), Argyre (1300 km), Isidis (1500 km), Ladon (1000 km). The MBA-based MPF is shown in blue (see text; also provided in the Supplemental Material along with a NEO-based MPF).*

## 6. Conclusions

Here we have presented a new crater chronology for Mars based on the latest dynamical models. We stress that because martian chronologies ultimately derive from lunar chronologies, it is important to disentangle the various impactor populations. While uncertainties remain in the earliest evolution of the Solar System, our chronology significantly differs from previous ones.

In particular, using a published cratering data (Shahrzad et al 2019), we estimate an age for the dark-toned units in Jezero crater to be ~ 3.1 Ga, or up to 0.5 Ga older than previously thought. We stress, however, that this age is strongly dependent on both the impactor and measured crater SFDs. Therefore, the determination of a radiometric age of the dark-toned unit in Jezero crater could provide a valuable constraint for the new Mars crater chronology and remove residual uncertainties due to crater scaling laws and or impactor SFDs.

We also show that all martian chronologies overestimate the number of martian craters larger than 150 km. This may underline a fundamental problem of available cratering chronologies to be addressed in future work. Here, we note that this discrepancy can be reconciled by a combination of timing of the Borealis basin formation and large crater erasing. We show that a Borealis age of ~ 4.20 Ga would be required if $N_{150}$ crater erasing is solely due to Borealis basin formation. Such an age, however, is not compatible with the assumed impact flux. In addition, our inferred age for Isidis basin 4.05-4.20 Ga provides further support for a Borealis basin older than 4.2 Ga. We estimate that the youngest possible age for the Borealis basin to be 4.35-4.40 Ga. This age satisfies our impact flux model and the observed number of 150 craters, if 50% of the craters larger than 150 km that formed after the Borealis basin were erased, as discussed in previous sections. An older Borealis age would require even more large crater erasure, and the increased number of basins larger than 500 km would not be compatible with geophysical constraints from Borealis basin's rim. This discussion indicates that a solid radiometric age of the Borealis basin and or the Isidis basin would be a very valuable constraint for earliest martian chronology.

**Acknowledgment.** SM thanks B. Bottke, A. Morbidelli, D. Nesvorny, and S. Robbins for interesting discussions about Mars things. SM thanks C. Fassett for his accurate review and discussion that greatly improved the manuscript. SM acknowledges support from NASA Habitable Worlds grant NNX16AR87G.